\documentclass[aps,amssymb,showkeys,letterpaper]{revtex4}

\usepackage{array}
\usepackage{multirow}
\usepackage{amsmath}
\usepackage{amsthm}
\usepackage{graphicx} 
\usepackage{amssymb}
\usepackage[pdftex, bookmarks, colorlinks=true, plainpages = false, citecolor = red, urlcolor = blue, filecolor = blue]{hyperref}
\usepackage[usenames,dvipsnames]{color}
\usepackage{mathrsfs}
\usepackage{amscd}
\usepackage{rotating}

\newcommand{\be}{\begin{equation}}
\newcommand{\ee}{\end{equation}}
\newcommand{\bea}{\begin{eqnarray}}
\newcommand{\eea}{\end{eqnarray}}
\newcommand{\bean}{\begin{eqnarray*}}
\newcommand{\eean}{\end{eqnarray*}}

\theoremstyle{plain}

\theoremstyle{definition}

\allowdisplaybreaks

\begin{document}
\title{Percolation crossing probabilities in hexagons: a numerical study}

\date{\today}

\author{Steven M.\ Flores}
\email{steven.flores@helsinki.fi} 
\affiliation{
Department of Mathematics \& Statistics, \\ 
P.O. Box 68, 00014 University of Helsinki, Finland,\\
and\\
Department of Mathematics, University of Michigan, Ann Arbor, MI, 48109-2136, USA}

\author{Robert M.\ Ziff}
\email{rziff@umich.edu} 
\affiliation{Center for the Study of Complex Systems and Department of Chemical Engineering, University of Michigan, Ann Arbor, MI, 48109-2136, USA}

\author{Jacob J.\ H.\ Simmons}
\email{jacob.simmons@mma.edu}
\affiliation{Maine Maritime Academy, Pleasant Street, Castine, ME, 04420, USA}

\begin{abstract}  
In a recent article \cite{js}, the last author of this article used $c=0$ logarithmic conformal field theory to predict crossing probabilities for percolation clusters inside a hexagon with free boundary conditions.  In the present article, we verify these predictions with high-precision computer simulations for equiangular hexagons with side lengths alternating from short to long.  Our simulations generate percolation-cluster perimeters with hull walks on a triangular lattice inside a hexagon.  Each sample comprises two hull walks, and the order in which these walks strike the bottom and upper left/right sides of the hexagon determines the crossing configuration of the percolation sample.  We compare our numerical results with the predicted crossing probabilities, finding excellent agreement.
\end{abstract}

\maketitle

\section{Introduction}\label{intro}

\emph{Bernoulli percolation} \cite{sa,grim} on a triangular lattice (which we simply call ``percolation" in this article) is the assignment of one (the ``activated" or ``occupied" state) or zero (the ``deactivated" or ``vacant" state) to the lattice sites with respective probabilities $p$ and $1-p$ and independently of the states of the other lattice sites.  If the lattice is confined to a domain with a boundary, then the event that we can follow a path along nearest-neighboring occupied sites from one boundary segment to another nonadjacent boundary segment is a \emph{crossing event}, and its probability is a \emph{crossing probability}.  Crossing probabilities are natural observables that indicate the unique critical point $p_c\in(0,1)$ of the continuum limit of percolation.  By continuum limit, we mean the limit obtained through sending the lattice spacing $a$ to zero while simultaneously increasing the number of sites in the system so the triangular lattice always fills the original system domain.  If $p<p_c$ (resp.\ $p>p_c$), then the probability of a crossing event in the continuum limit is zero (resp.\ one), but if $p=p_c$, then this probability is strictly between zero and one.  All results in this article pertain to critical percolation: $p=p_c$.  (There are still interesting off-critical results on crossing probabilities in the continuum limit.  For example, G.\ Delfino and J.\ Viti \cite{dv} have determined the asymptotic behavior of off-critical rectangle crossing probability as a function of $x=a|p-p_c|^{-4/3}$ in a special version of the continuum limit, called the ``scaling limit," where we send $a\rightarrow0$ and $p\rightarrow p_c$ in such a way that $x$ remains fixed.)  Furthermore, crossing probabilities are invariant under conformal transformations of the system domain (M. Aizenman, see \cite{lps}).

\begin{figure}[t]
\centering
\resizebox{\textwidth}{!}{\includegraphics{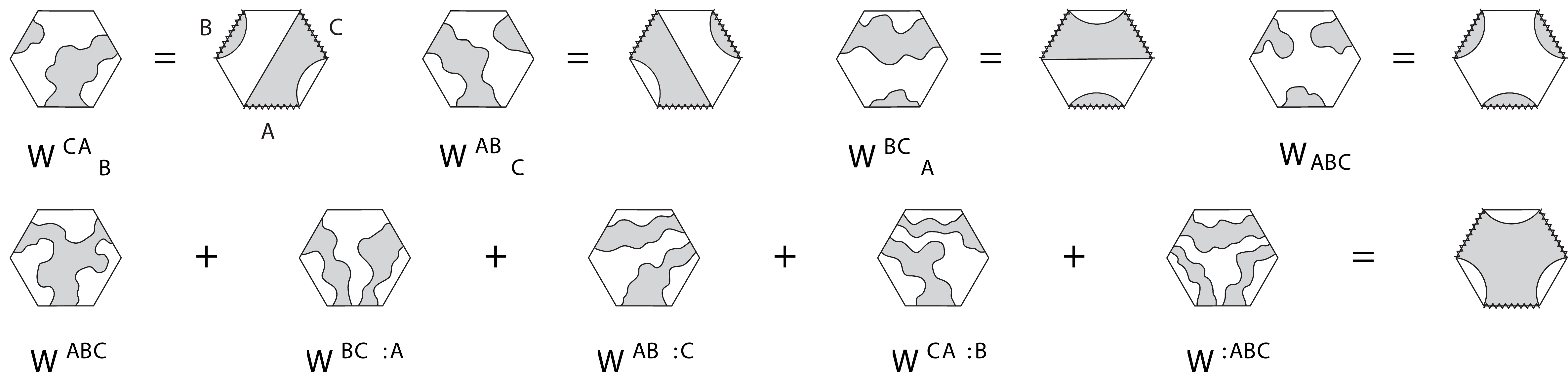}}
\caption{Percolation cluster-crossing configurations in a hexagon, with clusters filled gray.  The bottom, upper-left, and upper-right sides of each hexagon are labeled $A$, $B$, and $C$ respectively.  Configurations on the left (resp.\ right) sides of these graphical equations have the free BC (resp.\ their wired sides indicated by zig-zags).  The text below fully explains our notation.}
\label{Xings}
\end{figure}

As indicators of a conformally invariant critical point, crossing probabilities have been the subject of active research for some time (e.g., \cite{dv,lps,rsk,yuge,ziff92,pm}).  R.\ Langlands et al.\  verified their conformal invariance in various domains and on various lattices via computer simulations \cite{lps, lpps}.  Assuming that critical percolation scales onto a $c=0$ conformal field theory (CFT) in the continuum limit  \cite{bpz, fms}, J.\ Cardy correctly predicted the probability, given by \emph{Cardy's formula}, that a percolation cluster touches both opposite sides of a rectangle \cite{c3}.  (Technically, Cardy's result pertains to bond percolation, in which bonds between lattice sites are activated or deactivated in place of the lattice sites themselves.  However, strong numerical evidence and heuristic arguments support the notion of \emph{universality} \cite{lpps}.  This term means that in the continuum limit and exactly at the critical point, observables of Bernoulli percolation on a regular lattice are independent of microscopic details such as the the shape of the lattice and if lattice sites are activated or if, instead, bonds between them are activated.  For this reason, Cardy's results apply to Bernoulli site percolation on the triangular lattice, the model that we exclusively consider in this article.)  Using similar CFT methods, G.\ Watts correctly predicted the probability, given by \emph{Watts's formula}, that a percolation cluster touches all four sides of a rectangle \cite{watts}.  Because methods based on CFT are non-rigorous, these results were essentially well-motivated predictions verified to high accuracy \cite{ziff}.  More recently, Schramm-L\"owner evolution (SLE$_\kappa$) \cite{rohshr, knk, lawler} has delivered rigorous proofs for many of these predictions and produced new results.  A highlight of these findings is S.\ Smirnov's proof of Cardy's formula for the continuum limit of critical percolation on the triangular lattice via SLE$_6$ \cite{smirnov}. Subsequently, J.\ Dub\'edat \cite{dub} and later, O.\ Schramm, discovered proofs of Watts's formula via SLE$_6$ \cite{shefwilson}.  Further results include Schramm's formula \cite{schr}, crossing probabilities for related lattice models, each with a conformally invariant critical point, in rectangles \cite{bbk}, crossing probabilities for those models inside polygons with conformally invariant boundary conditions (BC) \cite{dub2, js, fksz}, crossing-cluster densities in rectangles \cite{skfz}, and probabilities of other percolation crossing events in rectangles \cite{skz}.

If we sample critical percolation inside a hexagon with the \emph{free} BC (i.e., we do not condition the state of any lattice site within a side of the hexagon), then in each sample, various percolation clusters may join together the hexagon's bottom side $A$ with its upper-left side $B$ and/or upper-right side $C$ in some crossing configuration. There are nine such fundamental crossing configurations whose collective probabilities sum to one, and they appear exclusively on the left sides of the graphical equations in figure \ref{Xings}.  Beneath each, we label their respective probabilities as $W^{ABC}$, $W^{AB}_{\hphantom{AB}C}$, $W^{AB:C}$, $W_{ABC}$, etc.

The notation of these probabilities requires explanation.  A superscript ``$X$" indicates that a percolation cluster connects side $X\in\{A,B,C\}$ of the hexagon with another side in this set, and a subscript ``$X$'' indicates that side $X$ is isolated from all other labeled sides.  Also, the notation ``$BC$:$A$" indicates that a percolation cluster connects side $A$ with side $B$, a different cluster connects side $A$ with $C$, but no cluster connects $B$ with $C$.  The side whose label follows the colon connects to the sides preceding the colon through two distinct clusters.  This type of double connection is related to the CFT boundary stress tensor \cite{js}, although this detail is unimportant to this article. Finally, the notation ``:$ABC$" indicates that a percolation cluster connects $A$ with $B$, a different cluster connects $B$ with $C$, and yet a different cluster connects $C$ with $A$.  In other words, $A$, $B$, and $C$ all touch two distinct crossing clusters, while no one cluster touches all three of these sides (figure \ref{Xings}).

In this article, we measure via computer simulation the probabilities of all of the nine critical percolation crossing events appearing on the left sides of the graphical equations in figure \ref{Xings}.  We exclusively sample various equiangular hexagons with the free BC on all sides and with side lengths that alternate between two values.  This restriction on the shape of the hexagon reduces the number of independent crossing probabilities from nine to five.
 
The free BC, exclusively considered in this article, differs from the BC that naturally arises when we use CFT \cite{bpz,c3} or multiple-SLE$_6$ \cite{bbk, dub3} to calculate percolation crossing probabilities.  In the latter situation, the natural BC to consider alternates from \emph{fixed}, or \emph{wired} (all sites occupied), to free as we pass from one side of the hexagon to the next, and we call this the \emph{free/fixed side-alternating BC} (FFBC) \cite{fksz}.  Because neighboring sites do not interact in percolation (unlike in other conformally invariant critical lattice models such as the Potts model \cite{wu}), a fixed side does not influence the states of the interior sites, so distinguishing between the free BC and the FFBC seems immaterial. However, changing the BC of a side from fixed to free may change the crossing configuration of a sample.  For example, we consider samples that contribute to $W^{BC:A}$, which have no spanning cluster between sides $B$ and $C$ but do have disjoint clusters connecting both of these sides to side $A$.  Upon wiring (i.e., activating all sites in) $A$, we create a crossing path from side $B$ to side $C$ that passes through both disjoint clusters and necessarily through side $A$ \cite{skz,shefwilson}.  In other words, by wiring side $A$, we create a crossing path between sides $B$ and $C$ in a sample where no such crossing path exists with the free BC.  From this observation, it is easy to see that percolation crossing probabilities for a hexagon with the free BC are building blocks of percolation crossing probabilities for a hexagon with the FFBC.  In pursuit of the former, it is useful to consider the latter.

\begin{figure}[t]
\centering
\resizebox{.7\textwidth}{!}{\includegraphics{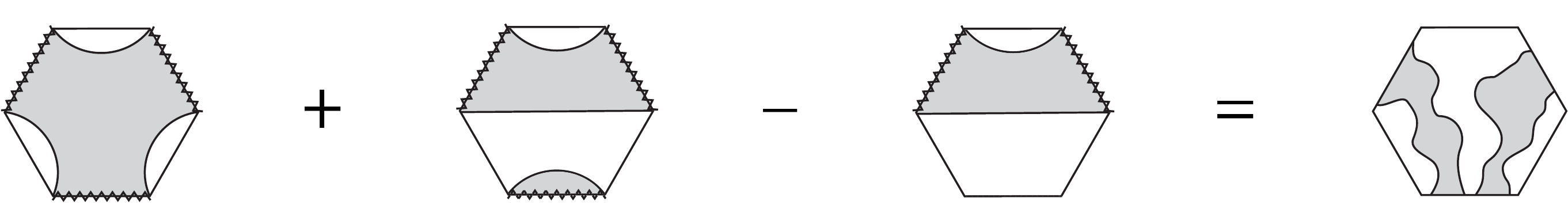}}
\caption{After subtracting the rectangle crossing probability from the total probability of the two FFBC hexagon crossing configurations on the left, we find the probability of the free-BC hexagon crossing configuration on the right.  With the bottom side of the ``subtracted" hexagon free, the gray percolation cluster above it may or may not touch that bottom side.}
\label{Decomp}
\end{figure}

There are five distinct percolation crossing configurations in which clusters join together the wired sides $A$, $B$, and $C$ of a hexagon exhibiting the FFBC \cite{fksz}, and \cite{dub2, js, fksz} give their probabilities.  Illustrations of these configurations appear on the right sides of the graphical equations in figure \ref{Xings}.  Moreover, and as previously noted, we may partition the sample space of percolation in the same hexagon but with the free BC into nine distinct cluster-crossing configurations that involve sides $A$, $B$, and $C$.  If we remove the BC distinction by eliminating all of the sites along the perimeter of the hexagon, then four crossing events of the free-BC sample space correspond one-to-one with four crossing events of the FFBC sample space, whose probabilities we already know (top line of figure \ref{Xings}).  Next, the remaining five crossing events of the free-BC sample space combine to give the remaining event of the FFBC sample space, whose probability we already know (bottom line of figure \ref{Xings}).  Figure \ref{Decomp} shows how we may use Cardy's formula \cite{c3} to find three, $W^{BC:A}$, $W^{CA:B}$, and $W^{AB:C}$, of those remaining five free-BC probabilities. However, the probabilities $W^{ABC}$ and $W^{:ABC}$ remain combined.  Recently, one of the authors used $c=0$ logarithmic CFT to isolate these last two probabilities \cite{js}, thereby predicting explicit formulas for all nine free-BC hexagon percolation crossing probabilities appearing on the left sides of the equations in figure \ref{Xings}.  

In this article, we measure via computer simulation the probabilities of the nine critical percolation crossing events in the free-BC hexagon.  As we explain in section \ref{Methodology}, three numbers, $0<m_1<m_2<m_3<1$, determine a unique equiangular hexagon up to size, position, and orientation, so thanks to their conformal invariance, all crossing probability formulas depend exclusively on these three numbers.  Although one must sample this entire three-dimensional parameter space  to completely verify these formulas, we sample only a one-dimensional subspace of it, that pertaining to equiangular hexagons with side lengths alternating from short to long, for practical reasons.  (Besides hexagons, one may consider more general ``conformal hexagons," or Jordan domains with six marked boundary points.  This includes polygons with any number of sides and with the marked points at different locations, perhaps but not necessarily at vertices, on the perimeters of these shapes.  Although perhaps interesting, it is not necessary to test these shapes upon verifying the crossing probability formulas for the entire three-dimensional parameter space of the equiangular hexagon.  Indeed, conformal invariance of critical percolation \cite{lps,smirnov} combined with the Riemann mapping theorem extends this verification to all conformal hexagons.)  In section \ref{Methodology}, we explain the method of our simulations and compare our measurements with the formulas predicted in \cite{js}, finding excellent agreement.

\section{Methodology and Results}\label{Methodology}

\begin{figure}[p]
\centering
\includegraphics[scale=0.25]{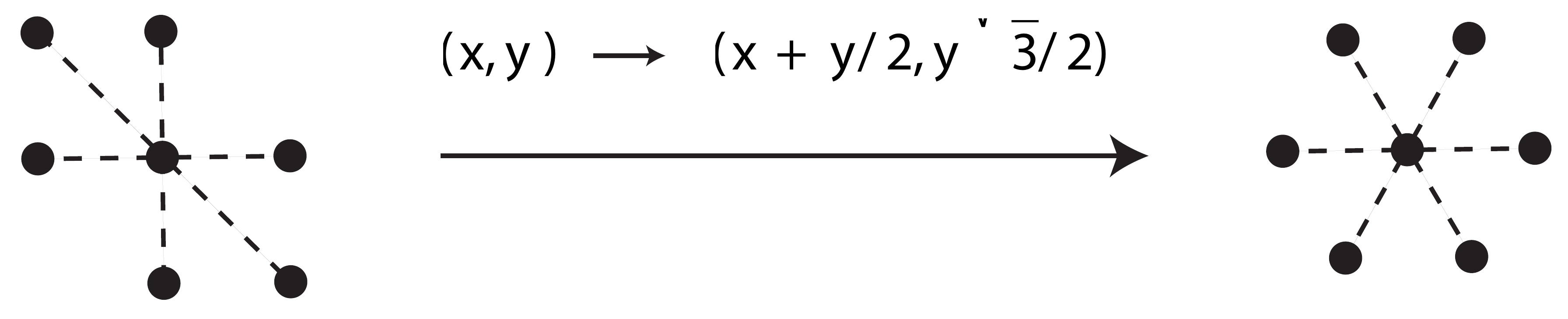}
\caption{Transformation from the square lattice to the triangular lattice.  The dashed lines connect the center site with its nearest neighbors on either lattice.}
\label{TriLat}

\resizebox{\textwidth}{!}{\includegraphics{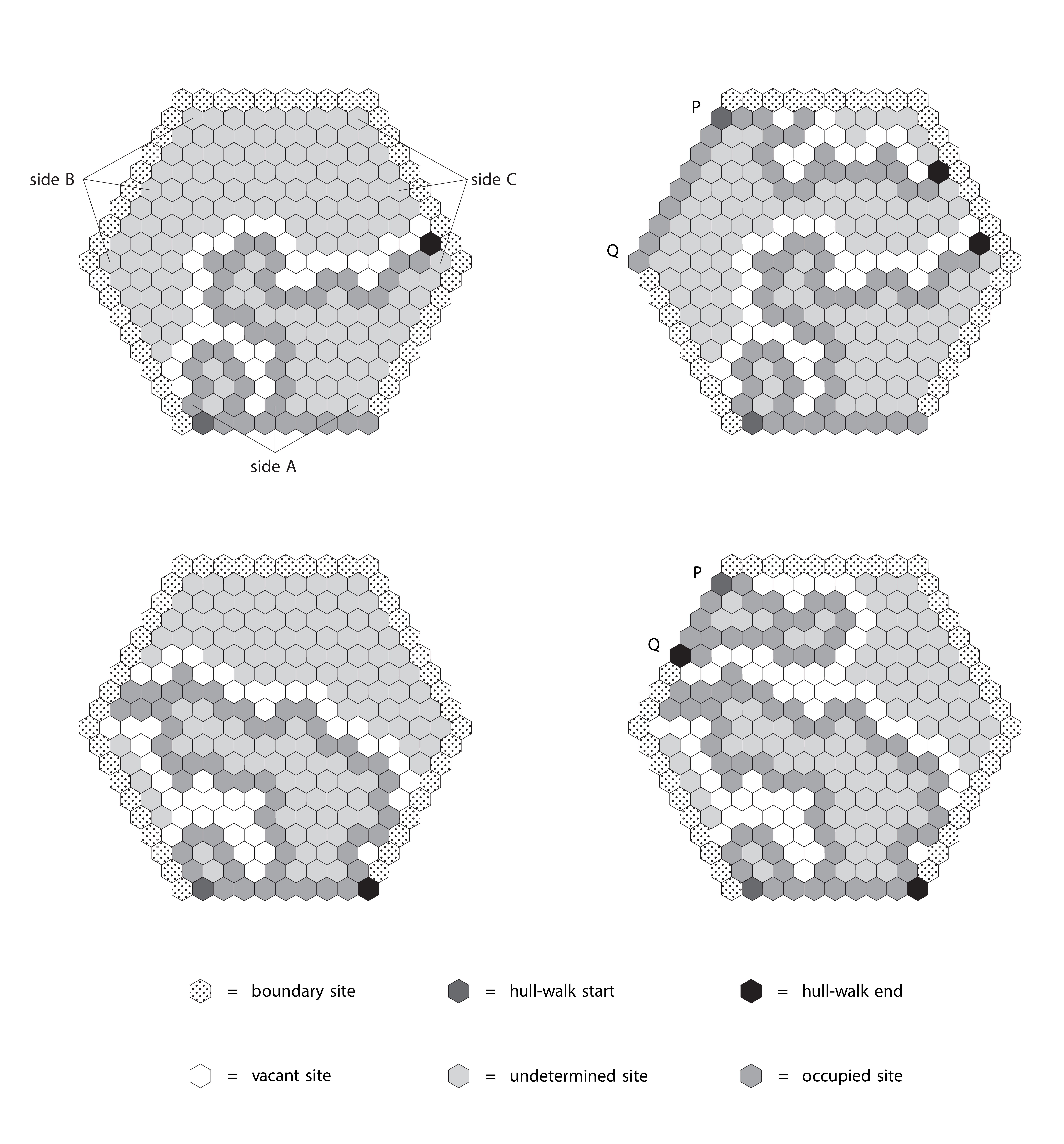}}
\caption{Each left hexagon shows a completed first hull walk, and each right hexagon shows a completed second hull walk after the first hull walk ends in the left hexagon.  The top-right (resp.\ bottom-right) hexagon contributes to $W^{AB:C}$ (resp.\ $W^{AB}_{\hspace{.43cm} C}$).}
\label{HexHullWalk}
\end{figure}

In this section, we describe how we sampled and measured the probabilities of the nine free-BC hexagon percolation cluster-crossing events described in section \ref{intro}.  Our simulations sampled critical site percolation on a triangular lattice (created from a square lattice via the transformation in figure \ref{TriLat}) inside an equiangular hexagon.  The critical activation probability for site percolation on the triangular lattice is $p_c=1/2$ \cite{smirnov}.  We considered only hexagons with their side lengths alternating between two values $\ell$ and $\ell'.$  Our simulations tested thirty three hexagons with different side-length ratios $R:=\ell/\ell'$ and generated 3,276,800 samples for each hexagon.  Each hexagon had about 3,000,000 sites (table \ref{SummaryTable}).  The time required to generate all samples for one hexagon using a 2GHz processor was about seventeen hours.

To create the equiangular hexagon described above, we mapped the upper half-plane onto the hexagon's interior via the Schwarz-Christoffel transformation
\be\label{ftrans}f(z)=\frac{2}{3}\sideset{}{_0^z}\int\zeta^{-1/3}(m_1-\zeta)^{-1/3}(m_2-\zeta)^{-1/3}(m_3-\zeta)^{-1/3}(1-\zeta)^{-1/3}\,d\zeta.\ee
This map (\ref{ftrans}) sends $0<m_1<m_2<m_3<1<\infty$ counterclockwise to the respective vertices $w_1,$ $w_2,\ldots,w_6$ of the hexagon.  In particular, $w_1=f(0)=0$, $w_2=f(m_1)>0$, and $w_3=f(m_2)\in\mathbb{H}$, so the hexagon lives in the upper half-plane, with its base flush against the real axis.  Furthermore, we let
\be\label{m1m3}m_1=\frac{m_2^2}{1-m_2+m_2^2},\quad m_3=\frac{m_2}{1-m_2+m_2^2},\ee
so the side lengths of the hexagon alternate from long to short.  Hence, if $\ell$ (resp. $\ell'$) is the length of the bottom (resp.\ top) side, then
\be\label{ratioside}\ell=f(m_1),\quad \ell'=|f(m_2)-f(m_1)|\quad\Longrightarrow\quad R=\frac{f(m_1)}{|f(m_2)-f(m_1)|}.\ee
Throughout this article, we refer to the bottom side, the upper-left side, and the upper-right side of the hexagon as side $A$, side $B$, and side $C$ respectively, and all of these sides have length $\ell$.

Relations (\ref{ftrans}--\ref{ratioside}) put $R\in(0,\infty)$ in one-to-one correspondence with $m_2\in(0,1)$, so $m_2$ completely determines the shape of the hexagon.  We can manipulate these relations using the symmetry of the hexagon to find the following expression for the aspect ratio in terms of $m_2$:
\be\label{explicitR}
R=\dfrac{\displaystyle{\int_0^1} \left[\zeta(1-\zeta)\left(1-\frac{m_2^2}{1-m_2+m_2^2}\zeta\right)(1-m_2 \zeta)\left(1-\frac{m_2}{1-m_2+m_2^2}\zeta\right)\right]^{-1/3}\,d\zeta}{\displaystyle{\int_0^1} \left[\zeta(1-\zeta)\left(1-\left(1-\frac{m_2}{1-m_2+m_2^2}\right)\zeta\right)(1-(1-m_2) \zeta)\left(1-\left(1-\frac{m_2^2}{1-m_2+m_2^2}\right)\zeta\right)\right]^{-1/3}\,d\zeta}.
\ee

Instead of generating entire percolation samples, our simulations only generated perimeters of percolation clusters anchored to sides $A$ and $B$ with two hull walks on the triangular lattice \cite{grass,z,wt} (figure \ref{HexHullWalk}).  These walks gave sufficient information to determine the crossing configuration of a generated percolation sample.  Our simulations proceeded as follows.  After surrounding the hexagon with two nested rings of lattice sites, which we call the ``inner ring" and the ``outer ring" (figure \ref{HexHullWalk} shows only the inner ring), we initialize all sites in the hexagon to an ``undetermined" state (until such a time at which the hull walk may visit the sites to determine their state).  Then we set the state of all sites in the two nested rings to ``boundary," meaning that a hull walk could not step on those sites.  (Later, we changed the states of some sites in the inner ring so the hull walks could conveniently step on them.)  To determine the crossing configuration of a sample, we tracked the order in which the hull walks stepped on sides $A$, $B$, and $C$.  Then we generated new hull walks in the same way as was just described to create the next sample.

Starting at the leftmost site in the part of the inner ring beneath side $A$, the first walk generated the perimeters of the leftmost (and sometimes all of the) percolation clusters anchored to side $A$.  In order to generate all of these disjoint perimeters with one walk, we changed the state of all sites in the inner ring and beneath side $A$ from ``boundary" to ``activated" before the walk started.  Doing this allowed the walk to travel from one cluster anchored to side $A$ rightward to the next by stepping through the inner ring.  The sites beneath side $A$ were the only sites of the inner ring onto which the first hull walk could step (figure \ref{HexHullWalk}).  If the first walk stepped on side $C$ and turned clockwise, then it could not step on side $B$ again.  In this event, we had sufficient information to determine the crossing configurations of the clusters anchored to side $A$, so we ended the walk there (upper hexagons of figure \ref{HexHullWalk}).  Otherwise, the walk ended at the rightmost site in the part of the inner ring beneath side $A$ (lower hexagons of figure \ref{HexHullWalk}).

The second hull walk determined if a cluster connected sides $B$ and $C$ in the generated sample.  If the first walk stepped on the top side of the hexagon, then it blocked such a crossing cluster, so it was not necessary to perform the second walk.  Otherwise, before starting the second walk, we activated all sites in the inner ring that were behind side $B$ and between sites $P$ and $Q$ inclusive, where $P$ was immediately left of the hexagon's upper-left vertex and $Q$ was directly above the first walk's highest step on side $B$ (figure \ref{HexHullWalk}).  (If the first walk did not step on side $B$, then $Q$ was the site immediately left of the hexagon's middle-left vertex instead.)  Then starting at site $P$, the second walk generated the perimeters of the upper (and sometimes all of the) percolation clusters anchored to side $B$ between sites $P$ and $Q$.  If the walk stepped on side $C$, then a percolation cluster joined $B$ with $C$.  In this event, we had sufficient information to determine the crossing configuration of the sample, so we ended the walk there (figure \ref{HexHullWalk}).  Otherwise, the second walk ended at $Q$.

\begin{figure}[p]
\centering
\includegraphics[scale=0.37]{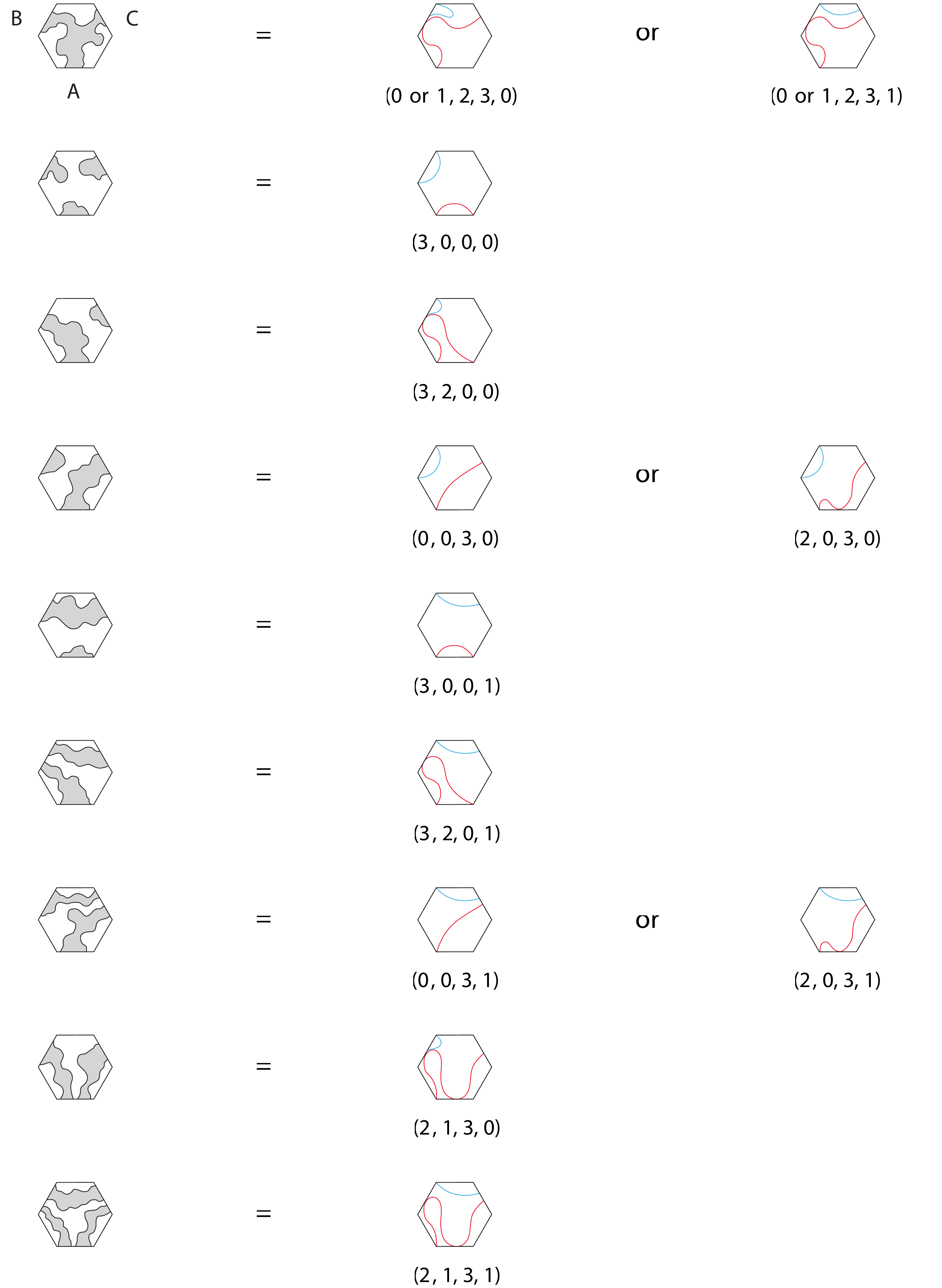}
\caption{Percolation-cluster crossings and hull-walk correspondence.  The first three numbers of each label give the hitting order of the first walk (red) against sides $A$, $B$, and $C$, and the last indicates if the second walk (blue) strikes side $C$.}
\label{Logic}
\end{figure}

After both walks finished, we determined the crossing configuration of the sample from the order in which the first walk stepped on sides $A$, $B$, and $C$ and whether or not the second walk stepped on side $C$.  The former determined how the percolation clusters anchored to side $A$ connected with sides $B$ and $C$. Three flag variables $a$, $b$, and $c$, each assuming different integer values in $\{0,1,2,3\}$, tracked this order.  Initialized to zero, we set $a=3$ immediately after the first walk stepped on the part of the inner ring just beneath side $A$, indicating that the walk had jumped from one percolation cluster anchored to side $A$ to the next one on the right.  Also initialized to zero, we set $b=3$ (resp.\ $c=3$) immediately after the first walk stepped on side $B$ (resp.\ $C$), indicating that the associated percolation cluster touched that side.  To track the orders of these events, we decreased by one all positive-valued flags $y\in\{a,b,c\}\setminus\{x\}$ immediately after setting $x=3$, where $x\in\{a,b,c\}$.  A fourth flag variable $d$, independent of the other three, equaled one if the second walk stepped on side $C$ and zero otherwise.  The ordered collection $(a,b,c,d)$ uniquely determined the crossing configuration of the sample.  Figure \ref{Logic} shows all possible values of $(a,b,c,d)$ and their corresponding crossing configurations.  Dividing the number of times that each of the nine possible crossing configurations occurred by the total number of samples gave measured values for the probabilities of the nine possible crossing events.

We generated the percolation hull walks described above in thirty three equiangular hexagon shapes, and in each, the long side together with the short side contained 2048 lattice sites in total.  The length $\ell_i$ (resp.\ $\ell_i'$) of (i.e., number of lattice sites in) the bottom (i.e., $A$), upper-left (i.e., $B$), and upper-right (i.e., $C$) (resp.\ top, lower-left, and lower right) sides of the $i$th hexagon was
\be\label{ell}\ell_i=\left\lfloor 120.471+56.4706(i-1)\right\rfloor,\quad\ell_i':=2048-\ell_i\approx\ell_{34-i},\quad i\in\{1,2,\ldots,33\}.\ee
These side lengths (\ref{ell}) are almost uniformly distributed between $\ell_1$ and $\ell_{33}$.  Furthermore, with $R_i:=\ell_i/\ell_i'$ the ratio  (\ref{ratioside}) of the side lengths of the $i$th hexagon, we have $R_1\approx1/16,$ $R_{17}=1$, and $R_{33}\approx16$ (table \ref{SummaryTable}).

Because the hexagon side lengths alternate from long to short thanks to (\ref{ftrans}--\ref{ratioside}), we have the following symmetric relations among the nine possible percolation crossing probabilities (figure \ref{Xings}):
\be\label{symm} W^{BC}_{\hphantom{BC}A\hphantom{:}}(R)=W^{CA}_{\hphantom{CA}B\hphantom{:}}(R)=W^{AB}_{\hphantom{AB}C\hphantom{:}}(R),\qquad W^{AB:C}(R)=W^{CA:B}(R)=W^{BC:A}(R).\ee
Hence, we must only measure probabilities of five of the nine possible crossing events.  Ref.\ \cite{js} predicts formulas for these probabilities, and they are
\bea\label{first}W_{ABC\hphantom{:}}(R)&=&2G(1-m_2)+G(m_2)-1,\\
W^{AB}_{\hphantom{AB}C\hphantom{:}}(R)&=&1-G(m_2)-G(1-m_2),\\
\label{middle}W^{BC:A}(R)&=&G(m_2)-\Pi(m_2),\\
\label{middlelast}W^{ABC\hphantom{:}}(R)&=&G(1-m_2)-G(m_2)+3\Pi(m_2)-H_a(m_2)+H_b(m_2)-1,\\
\label{last}W^{:ABC}(R)&=&H_a(m_2)-H_b(m_2),\eea
with $R$ and $m_2$ related via (\ref{explicitR}), and with the functions $\Pi$, $G$, $H_a$, and $H_b$ given by (below, we changed some of the integration variables from what is given in \cite{js} to facilitate the numerical integration in Mathematica)
\bea\Pi(\mu)&:=&\frac{\Gamma(2/3)}{\Gamma(1/3)\Gamma(4/3)}\left(\frac{\mu^2}{1-\mu+\mu^2}\right)^{1/3}\,_2F_1\left(\frac{1}{3},\frac{2}{3};\frac{4}{3}\,\Bigg|\,\frac{\mu^2}{1-\mu+\mu^2}\right),\\
\nonumber G(\mu)&:=&\frac{\Gamma(2/3)^2}{\Gamma(1/3)^4}\mu^{2/3}(1-\mu)^{2/3}(1-\mu+\mu^2)^{1/3}\int_0^1\int_0^1dw_1\,dw_2\, (1-\mu w_1+\mu^2 w_1 w_2)^{4/3}\\
&\times&\Big[w_1w_2(1-\mu w_1)(1-\mu w_2)(1-w_1)(1-\mu+\mu^2w_1)(1-\mu+\mu^2w_2)(1-\mu w_2+\mu ^2w_2)^2\Big]^{-2/3},\\
\nonumber H_a(\mu)&:=&\frac{\sqrt{3}\,\Gamma(2/3)^2}{4\pi\,\Gamma(1/3)^4}\mu^{8/3}(1-\mu)^{7/3}\int_0^1\int_0^1\int_0^1\int_0^1dw_1\,dw_2\,dw_3\,dw_4\,\,w_2w_4(1-\mu+\mu w_4)(1-\mu+\mu^2w_4)\\
\nonumber&\times&\Big[\mu w_2(1-w_3)(1-\mu w_1+\mu^2w_1-\mu^2 w_1w_2)(1-\mu w_1+\mu^2w_1-\mu^2w_1w_2w_3)\Big]^{4/3}\Big[w_1(1-w_1)(1-w_2)\Big]^{-2/3}\\
\nonumber&\times&\Big[(1-w_2w_3)(1-\mu w_1)(1-\mu+\mu w_2)(1-\mu+\mu w_2w_3)(1-\mu+\mu^2-\mu^2w_2)(1-\mu+\mu^2-\mu^2w_2w_3)\Big]^{-2/3}\\
&\times&\Big[(1-\mu+\mu^2w_1w_4)((1-\mu)^2+\mu w_4+\mu w_2-\mu^2w_2)((1-\mu)^2+\mu w_4+\mu w_2w_3-\mu^2w_2w_3)\Big]^{-2},\\
\nonumber H_b(\mu)&:=&\frac{\sqrt{3}\,\Gamma(2/3)^2}{4\pi\,\Gamma(1/3)^4}\mu^4(1-\mu)\int_0^1\int_0^1\int_0^1dw_2\,dw_3\,dw_4\,\,(1-w_3)^{4/3}w_4(1-\mu+\mu w_4)w_2^{7/3}\\
\nonumber&\times&\Big[(1-\mu+\mu w_2)(1-w_2)(1-\mu+\mu w_2w_3)(1-w_2w_3)\Big]^{-2/3}\\
&\times&\Big[(1-\mu)^2+\mu w_4+w_2w_3\mu(1-\mu)\Big]^{-2}\Big[1-\mu(1-w_4)-(1-w_2)\mu(1-\mu)\Big]^{-2}.\eea
The function $\Pi$ is Cardy's formula \cite{c3} for the hexagon treated as a conformal rectangle with opposite sides $B$ and $C$.  There, the hexagon has wired boundary conditions on these two sides, a percolation cluster touching both of these wired sides, and the free BC on the rest of the its boundary, as in the third hexagon from the left in figure \ref{Decomp}.

\begin{table}[t]
\begin{tabular}{r@{\qquad}c@{\qquad}c@{\qquad}c@{\qquad}c@{\qquad}c}
&
$W_{ABC}$&
$W^{AB}_{\phantom{AB}C}$&
$W^{ABC}$&
$W^{BC:A}$&
$W^{:ABC}$\\
\hline
\\[-2ex]
$R \to 0$&
1&
$\dfrac{\Gamma(1/3)^4}{3 \Gamma(2/3)^5} R^2$&
$\dfrac{2\Gamma(1/3)^4 }{3 \Gamma(2/3)^5} R^3$&
$\dfrac{\Gamma(1/3)^{22}}{5\cdot 3^8 \Gamma(2/3)^{23}} R^8$&
$\dfrac{\Gamma(1/3)^{53}}{50\cdot 3^{18} \Gamma(2/3)^{55}} R^{18}$\\[2ex]
$R \to \infty$&
$\dfrac{2\Gamma(1/3)^4 }{3 \Gamma(2/3)^5} \dfrac{1}{R^3}$&
$\dfrac{\Gamma(1/3)^4 }{3 \Gamma(2/3)^5} \dfrac{1}{R^2}$&
1/2&
$\dfrac{\Gamma(1/3) }{\Gamma(2/3)^2} \dfrac{1}{R}$&
1/2
\end{tabular}
\caption{Asymptotic behaviors or limiting values of the formulas (\ref{first}--\ref{last}) for probabilities of the percolation crossing events  (figure \ref{Xings}) in the equiangular hexagon with alternating side lengths.  $R$ is the side-length ratio of the hexagon, given by (\ref{explicitR}).}
\label{AsymptotesTable}
\end{table}

Table \ref{AsymptotesTable} gives asymptotic behaviors and/or limiting values for the crossing probabilities (\ref{first}--\ref{last}) as either $R \to 0$ or $R \to \infty$.  As $R\to0$, the hexagon's shape approaches the shape of an equilateral triangle, with the bottom side and upper-left/right sides of the former contracting to form the vertices of the latter.  In this limit, each hexagon crossing event with a percolation cluster joining at least two contracting sides goes to the event that a percolation cluster joins at least two vertices of the equilateral triangle.  Because the probability of the latter event is zero, the probabilities of the former events must vanish as $R\rightarrow0$.  Thus, we have
\be\label{lim1}W^{AB}_{\phantom{AB}C},W^{ABC},W^{BC:A},W^{:ABC}\rightarrow0\quad\text{and}\quad W_{ABC}\rightarrow1\quad\text{as $R\rightarrow0$}.\ee 
Next, as $R\rightarrow\infty$, the hexagon's shape again approaches the shape of an equilateral triangle, but now with the top side and lower-left/right sides of the former contracting to form the vertices of the latter.  In this limit, each hexagon crossing event with a percolation cluster joining at least two non-contracting sides goes to the event that a percolation cluster joins at least two sides of the equilateral triangle.  Because any two sides of the triangle are connected via an infinitesimal percolation cluster near their common vertex, the probabilities of the hexagon crossing events with at least one of the non-contracting sides isolated from another such side must vanish as $R\rightarrow\infty$.  Thus, we have
\be\label{lim2}W_{ABC},W^{AB}_{\phantom{AB}C},W^{BC:A}\rightarrow0\quad\text{and}\quad W^{ABC},W^{:ABC}\rightarrow1/2\quad\text{as $R\rightarrow\infty$.}\ee
The last two limits in (\ref{lim2}) follow from the necessary condition that $W^{ABC}+W^{:ABC}\rightarrow1$ as $R\rightarrow\infty$ and a straightforward argument that exploits the self-duality of site percolation on the triangular lattice.  Table \ref{AsymptotesTable} gives or implies all of these limits (\ref{lim1}, \ref{lim2}).

We used Mathematica v9 to numerically integrate the formulas for the crossing probabilities (\ref{first}--\ref{last}), where $m_2\in(0,1)$ uniquely corresponded with one of the thirty three tested hexagons via (\ref{explicitR}).  For the numerical integrations, we used the `NIntegrate' function with `MaxRecursion $\rightarrow30$' for $G$, `MaxRecursion $\rightarrow40$' and `WorkingPrecision $\rightarrow13$' and `Method $\rightarrow\{\text{``GlobalAdaptive", ``MaxErrorIncreases"}\rightarrow1000\}$' for $H_a$, and `MaxRecursion$\rightarrow30$' and `WorkingPrecision$\rightarrow15$' for $H_b$.

\begin{table}[b]
\begin{tabular}{p{1.8cm}p{2.5cm}p{2.5cm}p{2.2cm}p{2.5cm}p{2.5cm}p{1.75cm}}
crossing & $\hphantom{-}$avg.\ rel.\ err.\ & max.\ rel.\ err.\ & err.\ std.\ dev.\ & $\hphantom{-}$avg.\ rel.\ err.\ & max.\ rel.\ err.\ & err.\ std.\ dev.\ \\
probability & $\hphantom{-}$all hexagons & all hexagons & all hexagons & $\hphantom{-}13\leq i\leq 33$ & $13\leq i\leq 33$ & $13\leq i\leq 33$ \\
\hline
$W_{ABC}$ & $-0.000170$ & 0.005698 & 0.00260 & $-0.000111$ & 0.005698 & 0.00259 \\ 
$W^{AB}_{\hspace{.4cm}C}$ & $\hphantom{-}0.001065$ & 0.010349 & 0.00225 & $\hphantom{-}0.000339$ & 0.003480 & 0.00104 \\
$W^{BC:A}$ & $\hphantom{-}0.054657$ & 1.000000 & 0.26093 & $-0.001180$ & 0.004789 & 0.00331 \\
$W^{ABC}$ & $\hphantom{-}0.001682$ & 0.023209 & 0.00564 & $\hphantom{-}0.000008$ & 0.001335 & 0.00057 \\
 $W^{:ABC}$ & $\hphantom{-}0.308505$ & 1.000000 & 0.45734 & $\hphantom{-}0.002215$ & 0.099720 & 0.03307
\end{tabular}
\caption{Relative error between measured and predicted percolation crossing probabilities appearing in figures \ref{FirstPlots} and \ref{SecondPlots}, averaged over all thirty three hexagons tested, and standard deviation of all errors from their respective averages.}
\label{ErrorTable}
\end{table}

\begin{figure}[p]
\centering
\includegraphics[scale=0.24]{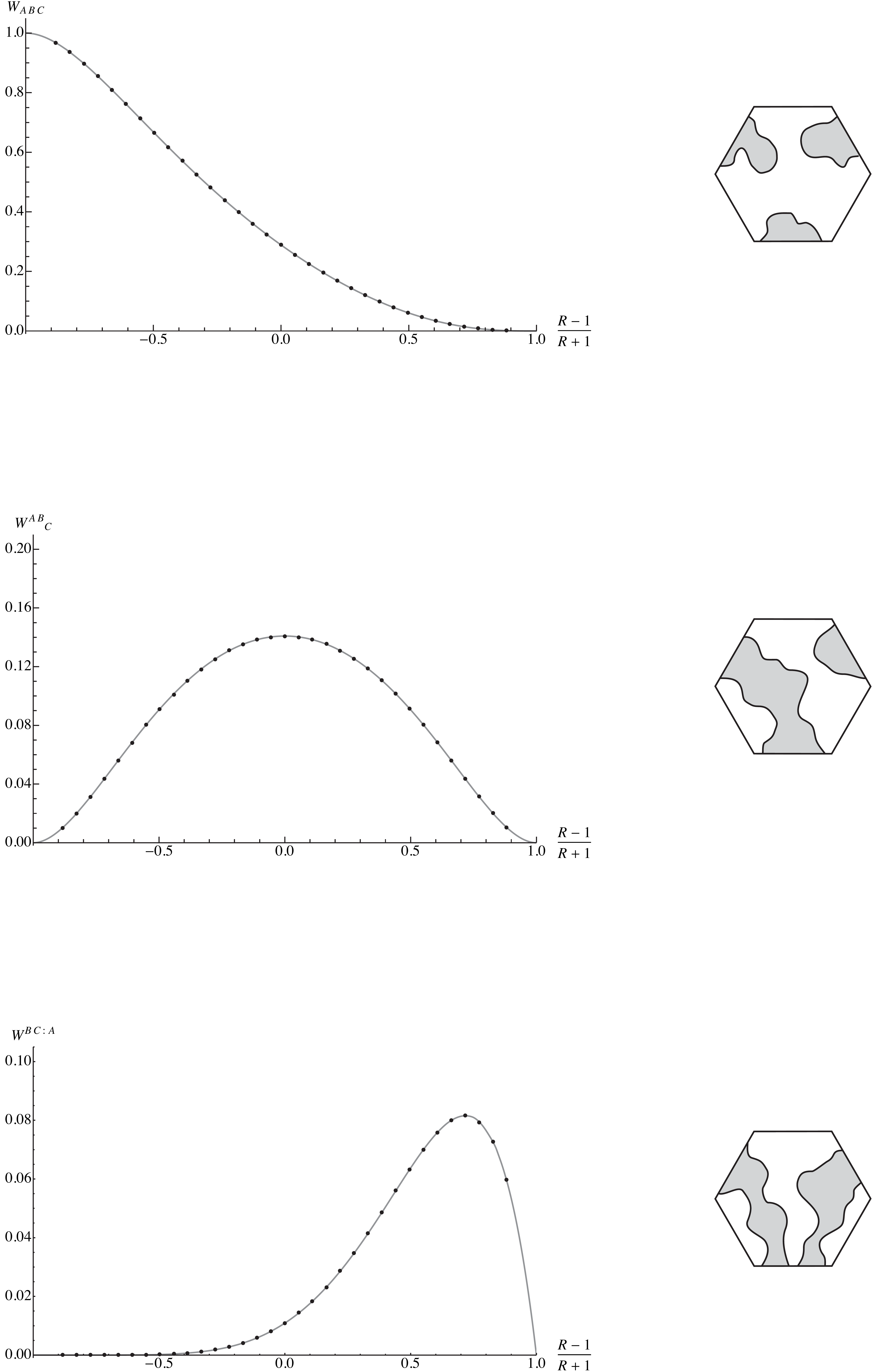}
\caption{Theory prediction and measurement of percolation cluster-crossing probabilities $W_{ABC}$, $W^{AB}_{\hspace{.4cm}C}$, and $W^{BC:A}$ (\ref{first}--\ref{middle}) versus $\frac{R-1}{R+1}$, with $R$ the ratio of the bottom side length to the top side length, in a hexagon with two alternating side lengths.
\label{FirstPlots}}
\end{figure}

We compare the theory predictions (\ref{first}--\ref{last}) with our measurements of the crossing probabilities in figures \ref{FirstPlots}--\ref{ErrorPlots} and in table \ref{SummaryTable}, noting that the errors are very small (on the order of about $10^{-4}$). In each figure, we plot all probabilities, measurements, and errors against the parameter $x=(R-1)/(R+1)\in[-1,1]$, with the limit $R\rightarrow0$ (resp. $R\rightarrow\infty$) corresponding to sending $x\rightarrow-1$ (resp.\ $x\rightarrow1$).  In the middle plot of figure \ref{FirstPlots}, we observe that the graph of $W^{AB}_{\phantom{AB}C}$ is symmetric about the line $x=0$.  This symmetry illustrates the general relation $W^{AB}_{\phantom{AB}C}(R)=W^{AB}_{\phantom{AB}C}(1/R)$ that arises from the self-duality of site percolation on the triangular lattice.

Table \ref{ErrorTable} gives the relative error between our predictions and our measurements in all of the five crossing events.  The second (resp.\ third) column of table \ref{ErrorTable} gives the average (resp.\ maximum) relative error between our predictions and measurements over all thirty three hexagons, and the fourth column gives the standard deviation of the relative errors from their average value.  The relative errors for the measurements of $W_{ABC}$, $W^{AB}_{\hspace{.4cm}C}$, and $W^{ABC}$ are very small, on the order of $10^{-3}$, indicating very good agreement between our measurements and the predicted probabilities.

On the other hand, the relative errors for the measurements of $W^{BC:A}$ and $W^{:ABC}$ in table \ref{ErrorTable} are not as small.  Indeed, our simulations generated no samples exhibiting their respective crossing events in the hexagons with smaller $R$ values (which justifies the corresponding maximum relative errors of one).  These latter crossing events are very rare if the bottom side of the hexagon is short compared to the top side (say, $R\leq0.5$) (table \ref{SummaryTable}), so in order to obtain better measurements of their probabilities, we must generate many more samples.  On the other hand, these same crossing events are not as rare if the lengths of the bottom and top sides of the hexagon are comparable (say, $R>0.5$), so our measurements should show much better agreement with our predictions for these hexagons.  Indeed, by restricting our attention to the thirteenth hexagon ($R_{13}=0.6384$) up to the thirty-third hexagon ($R_{33}=15.9256$) in the fifth, sixth, and seventh columns of table \ref{ErrorTable}, we find improved agreement.  In particular, the average relative errors are now on the order of $10^{-3}$.

\section{Summary}\label{Summary}

\begin{figure}[p]
\centering
\includegraphics[scale=0.24]{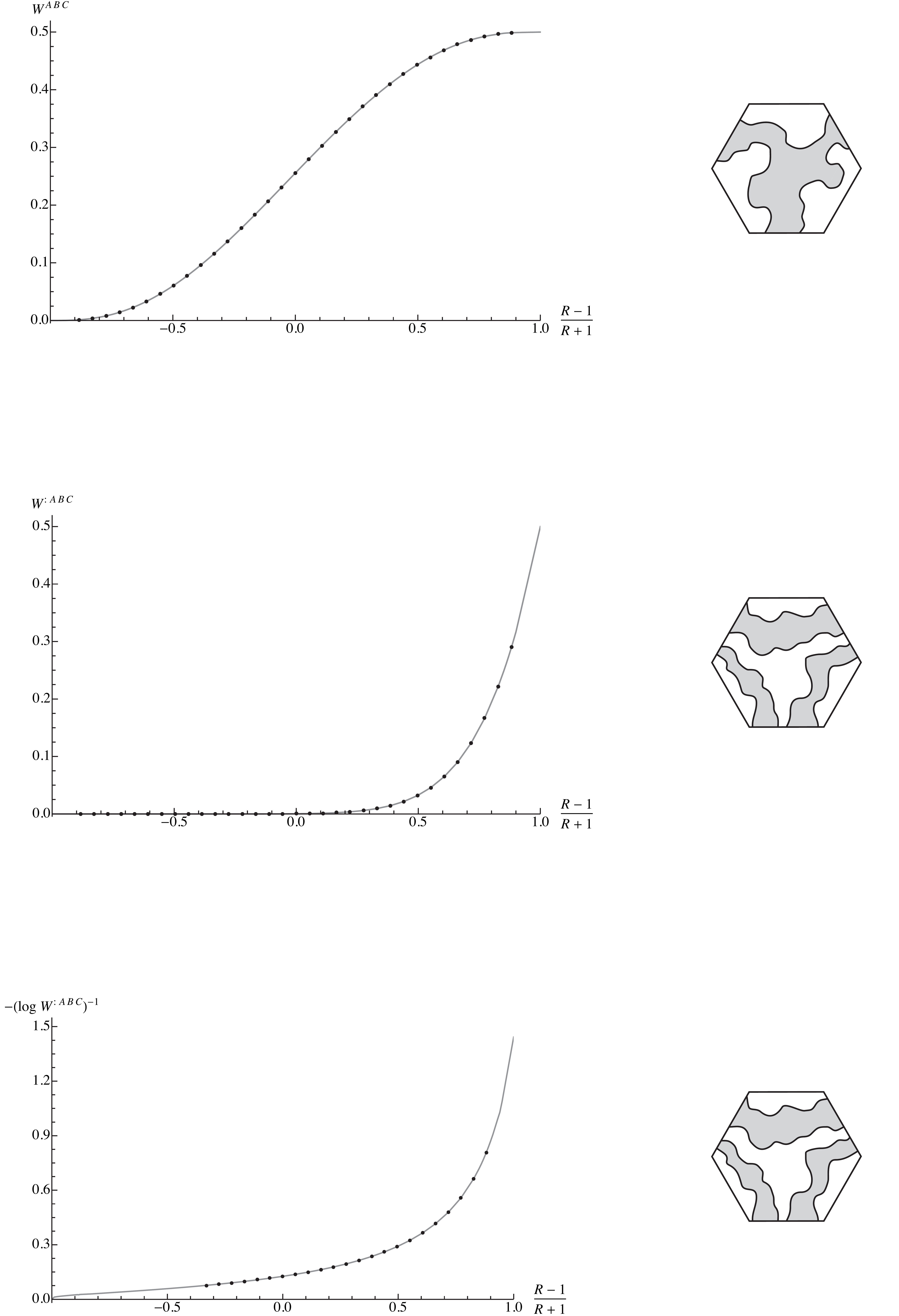}
\caption{Theory prediction and measurement of percolation cluster crossing probabilities $W^{ABC}$ and $W^{:ABC}$ (\ref{middlelast}--\ref{last}) versus $\frac{R-1}{R+1}$, with $R$ the ratio of the bottom side length to the top side length, in a hexagon with two alternating side lengths.  The bottom plot displays the information of the middle plot on a different scale that better visualizes the data and theory prediction for small $R$.  The bottom plot omits the first ten data points of the middle plot because our simulations did not generate any samples exhibiting the $W^{:ABC}$ crossing event for the corresponding values of $R$ (table \ref{SummaryTable}).}
\label{SecondPlots}
\end{figure}

\begin{figure}[p]
\centering
\includegraphics[scale=0.25]{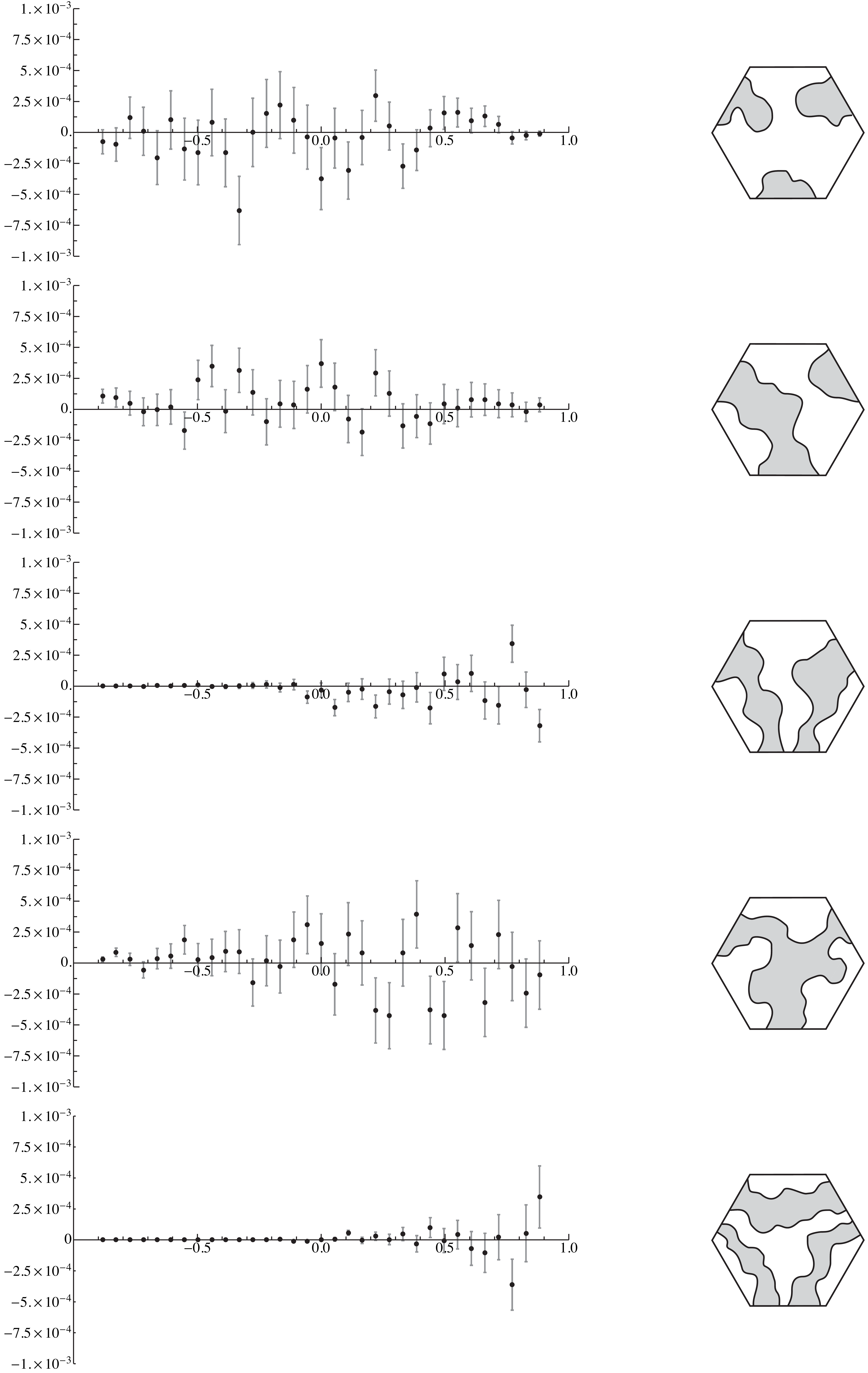}
\caption{Error plots for the five measured crossing probabilities (\ref{first}--\ref{last}).  Error bar widths are each one standard deviation $\sigma_i=\sqrt{W_i(1-W_i)/N}$, with $W_i$ the crossing probability for the $i$th hexagon and $N=3,276,800$ the number of samples.}
\label{ErrorPlots}
\end{figure}

In this article, we presented measurements of probabilities of all independent critical percolation crossing events that occur in an equiangular hexagon with the free BC and with side lengths alternating from long to short.  We modeled the continuum limit of percolation-cluster perimeters with hull walks on a large triangular lattice.  The order in which these hull walks struck the bottom and upper left/right sides of the hexagon determined the crossing configuration of the generated sample.  We compared our measurements with the predicted formulas (\ref{first}--\ref{last}) for these probabilities found in \cite{js} (figures \ref{FirstPlots}--\ref{ErrorPlots} and table \ref{SummaryTable}), finding excellent agreement with relative errors on the order of $10^{-3}$ (table \ref{ErrorTable}). Thus, we have obtained good numerical verification that the crossing-probability formulas derived in \cite{js} are correct for the hexagons described above.  These formulas add to the list of exact results for critical percolation and extend our understanding of this conformally invariant system.  

An obvious extension of this work is to test equiangular hexagons without constraints on the relative lengths of their sides.  One may obviously use the methodology of our simulations without modification to test these other hexagons.  We expect that results from such simulations would completely verify these hexagon percolation crossing-probability formulas.

\section{Acknowledgements}

The authors thank Peter Kleban for helpful comments and conversations and C.\ Townley Flores for proofreading the manuscript.

This work was supported by National Science Foundation Grants Nos.\,PHY--0855335 (SMF), DMR--0536927 (SMF), MRSEC DMR-0820054 (JJHS), DMR-0448820 (JJHS), DMR-0213745 (JJHS), and DMR-0906427 (JJHS).

\begin{sidewaystable}[p]
\centering
\renewcommand{\tabcolsep}{.2cm}
\begin{tabular}{cccccccccccccc}
\parbox{1.4cm}{hexagon number $i$} & \parbox{1.7cm}{bottom side length $\ell_i$} & $\parbox{1.7cm}{side-length ratio $R_i$ }$ & $\parbox{1.1cm}{number of sites }$ & $\parbox{1.1cm}{$W_{ABC}$ predict.\ }$ & $\parbox{1.1cm}{$W_{ABC}$ meas.\ }$ & $\parbox{1.1cm}{$W^{AB}_{\hphantom{AB}C}$ predict.\ }$ & $\parbox{1.1cm}{$W^{AB}_{\hphantom{AB}C}$ meas.\ }$ & $\parbox{1.1cm}{$W^{BC:A}$ predict.\ }$ & $\parbox{1.1cm}{$W^{BC:A}$ meas.\ }$ & $\parbox{1.1cm}{$W^{ABC}$ predict.\ }$ & $\parbox{1.1cm}{$W^{ABC}$ meas.\ }$ & $\parbox{1.1cm}{$W^{:ABC}$ predict.\ }$ & $\parbox{1.1cm}{$W^{:ABC}$ meas.\ }$ \\ [0.3cm]
\hline \\ [-0.3cm]
1	&	120	&	0.062241	&	2325441	&	0.967898	&	0.967974	&	0.010274	&	0.010168	&	0.000000	&	0.000000	&	0.001279	&	0.001249	&	0.000000	&	0.000000	\\
2	&	176	&	0.094017	&	2423553	&	0.936626	&	0.936724	&	0.019879	&	0.019784	&	0.000000	&
0.000000	&	0.003737	&	0.003650	&	0.000000	&	0.000000	\\
3	&	233	&	0.128375	&	2516976	&	0.898034	&	0.897915	&	0.031309	&	0.031260	&	 0.000000	&	             0.000001	&	0.008036	&	0.008006	&	     0.000000	&	0.000000	\\
4	&	289	&	0.164298	&	2602432	&	0.855515	&	0.855506	&	0.043407	&	0.043427	&	 0.000004	&	             0.000006	&	0.014251	&	0.014307	&	     0.000000	&	0.000000	\\
5	&	346	&	0.203290	&	2682973	&	0.809201	&	0.809405	&	0.056008	&	0.056013	&	0.000014	&	0.000010	&	0.022731	&	0.022695	&	     0.000000	&	0.000000	\\
6	&	402	&	0.244228	&	2755773	&	0.761982	&	0.761881	&	0.068229	&	0.068209	&	0.000039	&	0.000036	&	0.033215	&	0.033159	&	     0.000000	&	0.000000	\\
7	&	459	&	0.288861	&	2823432	&	0.713174	&	0.713309	&	0.080165	&	0.080337	&	0.000094	&	0.000090	&	0.046049	&	0.045861	&	     0.000000  &	0.000000	\\
8	&	515	&0.335943	&	2883576	&	0.665247	&	0.665409	&	0.091155	&	0.090918	&	0.000197	&	0.000186	&	0.060696	&	0.060671	&	     0.000000 &	0.000000	\\
9	&	572	&	0.387534	&	2938353	&	0.617073	&	0.616993	&	0.101411	&	0.101061	&	0.000383	&	0.000384	&	0.077547	&	0.077502	&	     0.000001	&	0.000000	\\
10	&	628	&	0.442254	&	2985841	&	0.570763	&	0.570928	&	0.110450	&	0.110465	&	0.000681	&	0.000685	&	0.095844	&	0.095751	&	     0.000001	&	0.000000	\\
11	&	685	&	0.502568	&	3027736	&	0.524976	&	0.525607	&	0.118511	&	0.118196	&	0.001150	&	0.001150	&	0.116038	&	0.115946	&	0.000002	&	         0.000002	\\
12	&	741	&	0.566947	&	3062568	&	0.481531	&	0.481531	&	0.125258	&	0.125121	&	0.001829	&	0.001823	&	0.137204	&	0.137361	&	0.000006	&	          0.000006 	\\
13	&	798	&	0.638400	&	3091581	&	0.439020	&	0.438867	&	0.130899	&	0.131000	&	0.002803	&	0.002790	&	0.159856	&	0.159838	&	0.000016	&	0.000015	\\
14	&	854	&	0.715243	&	3113757	&	0.399023	&	0.398802	&	0.135219	&	0.135175	&	0.004108	&	0.004119	&	0.182956	&	0.182985	&	0.000039	&	0.000035	\\
15	&	911	&	0.801231	&	3129888	&	0.360161	&	0.360063	&	0.138365	&	0.138329	&	0.005861	&	0.005848	&	0.207075	&	0.206886	&	0.000088	&	0.000098	\\
16	&	967	&	0.894542	&	3139408	&	0.323822	&	0.323860	&	0.140222	&	0.140060	&	0.008071	&	0.008160	&	0.231117	&	0.230808	&	0.000183	&	0.000193	\\
17	&	1024	&	1.000000	&	3142657	&	0.288717	&	0.289090	&	0.140855	&	0.140485	&	0.010888	&	0.010919	&	0.255690	&	0.255533	&	0.000365	&	0.000363	\\
18	&	1080	&	1.115700	&	3139521	&	0.256080	&	0.256126	&	0.140244	&	0.140063	&	0.014270	&	0.014443	&	0.279694	&	0.279866	&	0.000686	&	0.000681	\\
19	&	1136	&	1.245610	&	3130113	&	0.225278	&	0.225586	&	0.138409	&	0.138487	&	0.018309	&	0.018359	&	0.303335	&	0.303101	&	0.001235	&	0.001177	\\
20	&	1193	&	1.395320	&	3114096	&	0.195819	&	0.195859	&	0.135286	&	0.135470	&	0.023121	&	0.023145	&	0.326801	&	0.326719	&	0.002160	&	0.002164	\\
21	&	1249	&	1.563200	&	3092032	&	0.168749	&	0.168452	&	0.130987	&	0.130692	&	0.028539	&	0.028703	&	0.349060	&	0.349443	&	0.003614	&	0.003584	\\
22	&	1306	&	1.760110	&	3063133	&	0.143127	&	0.143076	&	0.125367	&	0.125240	&	0.034719	&	0.034763	&	0.370705	&	0.371130	&	0.005909	&	0.005905	\\
23	&	1362	&	1.985420	&	3028413	&	0.119888	&	0.120160	&	0.118642	&	0.118776	&	0.041364	&	0.041434	&	0.390788	&	0.390705	&	0.009307	&	0.009260	\\
24	&	1419	&	2.255960	&	2986632	&	0.098248	&	0.098391	&	0.110601	&	0.110657	&	0.048569	&	0.048578	&	0.409848	&	0.409455	&	0.014394	&	0.014425	\\
25	&	1475	&	2.574170	&	2939256	&	0.079020	&	0.078986	&	0.101581	&	0.101696	&	0.055872	&	0.056049	&	0.427052	&	0.427431	&	0.021570	&	0.021470	\\
26	&	1532	&	2.968990	&	2884593	&	0.061574	&	0.061417	&	0.091343	&	0.091301	&	0.063230	&	0.063131	&	0.442863	&	0.443286	&	0.031842	&	0.031847	\\
27	&	1588	&	3.452170	&	2824561	&	0.046578	&	0.046418	&	0.080368	&	0.080359	&	0.069987	&	0.069954	&	0.456597	&	0.456312	&	0.045756	&	0.045714	\\
28	&	1645	&	4.081890	&	2757016	&	0.033541	&	0.033445	&	0.068443	&	0.068366	&	0.075849	&	0.075745	&	0.468638	&	0.468499	&	0.064945	&	0.065014	\\
29	&	1701	&	4.902020	&	2684328	&	0.022943	&	0.022812	&	0.056229	&	0.056151	&	0.079939	&	0.080054	&	0.478494	&	0.478812	&	0.090059	&	0.090164	\\
30	&	1758	&	6.062070	&	2603901	&	0.014394	&	0.014330	&	0.043627	&	0.043582	&	0.081544	&	0.081698	&	0.486479	&	0.486248	&	0.123615	&	0.123593	\\
31	&	1814	&	7.752140	&	2518557	&	0.008132	&	0.008175	&	0.031520	&	0.031484	&	0.079566	&	0.079223	&	0.492355	&	0.492383	&	0.166257	&	0.166619	\\
32	&	1871	&	10.570600	&	2425248	&	0.003796	&	0.003821	&	0.020068	&	0.020088	&	0.072627	&	0.072656	&	0.496435	&	0.496678	&	0.221685	&	0.221632	\\
33	&	1927	&	15.925600	&	2327248	&	0.001309	&	0.001319	&	0.010426	&	0.010390	&	0.059423	&	0.059742	&	0.498802	&	0.498899	&	0.290342	&	0.289995	\\
\end{tabular}
\caption{The thirty three sampled hexagons with alternating side length, the length $\ell_i$ of (i.e., number of lattice sites in) the bottom side of, the side-length ratio $R_i=\ell_i/(2048-\ell_i)$ of, and number of lattice sites in the $i$th hexagon, and the predicted and measured values of the (only five thanks to (\ref{symm})) independent crossing probabilities for these hexagons (figures \ref{FirstPlots}, \ref{SecondPlots}).  In the rightmost column, the predicted values of $W^{:ABC}$ for $R\leq 0.5$ are very small but not exactly zero.  Indeed, their true values are comparable to the error of their numerical evaluation in Mathematica, so we show only the first six (vanishing) decimal places of their respective values.}
\label{SummaryTable}
\end{sidewaystable}

\end{document}